# Decentralized Data Governance as Part of a Data Mesh Platform: Concepts and Approaches

Arif Wider[1,2], Sumedha Verma[1], and Atif Akhtar[1]

[1]*Thoughtworks Germany*
Caffamacherreihe 7, 20355 Hamburg, Germany
{awider,sumedhav,syedatif}@thoughtworks.com

[2]*Hochschule für Technik und Wirtschaft Berlin*
Treskowallee 8, 10318 Berlin, Germany
wider@htw-berlin.de

*Abstract*—Data mesh is a socio-technical approach to decentralized analytics data management. To manage this decentralization efficiently, data mesh relies on automation provided by a self-service data infrastructure platform. A key aspect of this platform is to enable decentralized data governance. Because data mesh is a young approach, there is a lack of coherence in how data mesh concepts are interpreted in the industry, and almost no work on how a data mesh platform facilitates governance. This paper presents a conceptual model of key data mesh concepts and discusses different approaches to drive governance through platform means. The insights presented are drawn from concrete experiences of implementing a fully-functional data mesh platform that can be used as a reference on how to approach data mesh platform development.

## I. Introduction

Data mesh is a novel approach to analytics data management that was developed within Thoughtworks, a global IT consultancy, and was first presented in writing in 2019 [1]. It has since been described in detail in a seminal book on the topic [4]. Data mesh is an approach that is specifically addressing the problems of large IT organizations that try to create business value from their analytical data.

E-commerce scale-ups are among the first to adopt this new approach to improve the quality and effectiveness of their digital services. For example HelloFresh, a meal delivery platform, identified the challenge of data ownership and uses data mesh principles to drive clear goals and outcomes [15]. Zalando, an internet fashion platform, uses data mesh to enable frictionless communication and sharing between data consumers and producers [22]. An example from the financial sector is Saxo Bank that uses data mesh techniques to apply data governance in a decentralized ecosystem [7].

Such companies often face bottleneck situations with their central data systems and with their central data teams when scaling up the number of data sources and data consumption use-cases. Next to data quality issues, problems with scaling and distributing data responsibility have been stated to be the most common challenges and motivational factors to adopt a data mesh approach [17]. Data mesh sees the root cause of those issues in the centralization of data ownership. A central data team is often not familiar with the details of the data generation at its source system. They have to consult the team that maintains the data generating system every time something needs to be changed. As described in [2], this is a problem of an inefficient responsibility structure and thus cannot only be solved with technological improvements. Therefore, data mesh proposes a socio-technical solution by decentralizing data ownership to the *domain teams* that generate the data.

Overall, data mesh defines four principles for managing this decentralization efficiently:

- decentralized domain data ownership
- data as a product
- self-service data infrastructure platform
- federated computational data governance

Data as a product, the second principle, captures the idea of managing analytical data like a physical or digital product that you want to sell. Applying product management to a data set means that its development follows the demand of internal data customers. A data set managed that way is called a *data product*. It is noteworthy to point out that data products in the original data mesh concept are primarily about data exchange within one organization. This stands in some contrast to the existing literature on data economies and data marketplaces, where the term data product often has the connotation of data monetisation across organizations [18]. In this paper, however, we stick to the original meaning of the term in the context of data mesh, which is that a data product is an easy-to-consume and composable data offering that is used within an organization to create business value from data.

To give you an example of a data product within an example domain, let us consider a large e-commerce company similar to those described before; there would typically be many teams under several different departments, for example Finance, Marketing, Operations and so on. A *domain* in this case could just be the department itself, let's say "Marketing", and a data product in this domain could be "customer recommendations". We use the term domain according to its definition in the domain-driven design approach as described by Eric Evans in his seminal book on the topic [19]. Here, a domain refers to a *bounded context* of expertise and knowledge and consists of domain experts. A data product is a specific data offering that is designed to answer a specific set of questions such as "What product should we recommend to customer X?". It is built and maintained by a team inside a domain. In the remainder of this paper, we will use this imaginary e-commerce company for driving further examples of data products.

From an engineering perspective, data products apply the idea of information hiding to analytics data: a data product abstracts from implementation details and thereby allows people with less domain knowledge to get value from that data. Data products then serve as building blocks from which new higher-order data products can be composed. Data products provide clearly defined input and output ports to allow for this composition. A data mesh is a network of primitive and composed data products that are easily discoverable and consumable.

In this, data mesh resembles the microservices approach in general software engineering. Microservices is a modern reinterpretation of *service-oriented architecture* and most web services today are built using that architectural style. The success of the microservices architecture builds on the provision of powerful self-service infrastructure platforms, such as Amazon Web Services or Google Cloud Platform. They allow people with limited knowledge of computing hardware operations to effectively create and maintain large IT infrastructure. Similarly, in data mesh, the decentralization of data ownership to domain teams is not effective without a powerful self-service data infrastructure platform, which we from hereon call a *data mesh platform*.

The goal of a data mesh platform is to enable developers with limited data engineering knowledge to create, maintain, evolve, discover, compose, and ultimately decommission their data products. This means that a data mesh platform not only needs to automate the provision of data infrastructure, it also needs to provide an easy-to-use developer interface that ties all the different tools together. This interface of a data mesh platform is called a *data product developer experience plane*. A data mesh platform may provide other *planes*, i.e., interfaces at other abstraction levels. An example is a data infrastructure plane that allows for direct access to infrastructure automation.

Because data mesh is a young approach, there is a recurring need to be able to demonstrate how a data mesh works, what the key concepts of data mesh mean in practice, and what capabilities a data mesh platform needs to provide. For this reason, we have created a reference implementation of a data mesh platform that illustrates different approaches to common challenges such as data governance support. In this paper, we first present the conceptual model which is the basis of our reference implementation. We then dive into decentralized data governance and present selected approaches to it.

The remainder of the paper is organized as follows: Sect. II presents an extension of an existing model of key data mesh concepts. Next, in Sect. III, we look at the components of a data mesh platform and its required features. In Sect. IV, we define decentralized data governance and discuss different implementation approaches of how it can be supported by the platform. In Sect. V, we evaluate related work before concluding the paper in Sect. VI.

## II. A MODEL OF DATA PRODUCTS AND THEIR COMPOSITION

In his paper "Notation as a Tool of Thought" [24], Iverson argues that the notation or language that we use to discuss a problem will influence the solutions that we come up with. Similarly, the vocabulary we use in the data mesh domain give meaning and lend to the right (or the wrong) interpretation of data mesh. Therefore, in this section, we first present a selection of general data mesh concepts we deem most important for the topic of this paper. In later sections, we then relate those general data mesh concepts to specific data mesh platform concepts and to the requirements of data governance.

The conceptual model presented in [5] is a first attempt in providing a model of the main concepts of data mesh and their relations. We found, however, that it misses some key elements whose absence becomes especially apparent when implementing an enterprise-ready data mesh and its underlying data mesh platform. One of these omissions is that of input and output ports. Instead of providing an entirely new conceptual model, we decided to extend the model presented in [5] with those concepts that we believe provide the technical depth to achieve composability and scalability.

### A. How Input/Output Ports Facilitate Composability

Before we describe our extension, let us first illustrate why we deem the concept of a port so important for actual data mesh implementations. Fig. 1 shows a very small example data mesh consisting of only three data products. As illustrated by the dashed data mesh boundary, even in a greenfield data mesh setup not all data processing systems are part of the mesh. Instead, source data systems are represented in the mesh by so-called *source-aligned data products*.

Such source-aligned data products often have only one input port to consume data from the original source systems, but exhibit several output ports. The rationale behind providing several output ports is that the same data can be provided in different data formats or can be made available through different interfaces. For example, a data product which presents a useful abstraction over the original source data can provide its data in a streaming fashion via an Apache Kafka topic or in a distributed BLOB storage fashion via an AWS S3 bucket. Because the data product itself still manifests the same abstraction over the source data, it makes sense to have this abstraction encapsulated as one data product with two output ports and not as two separate data products.

Another category of data products are so-called consumer-aligned data products. Whereas source-aligned data products provide their data in a rather general way and in multiple formats in order to support many potential data consumption use-cases, a consumer-aligned data product is focused on one or only few specific data consumption use-cases that it optimizes for. The value of such a consumer-aligned data product is often to combine data from different source-aligned data products in a meaningful way. Consumer-aligned data products therefore often have several input ports but only one output port with a specific consumption use-case in mind. Our previously described "customer recommendations" data product is such a consumer-aligned data product which reads data from an upstream "customer tracking information" data product and a "customer details" data product, which are both source-aligned.

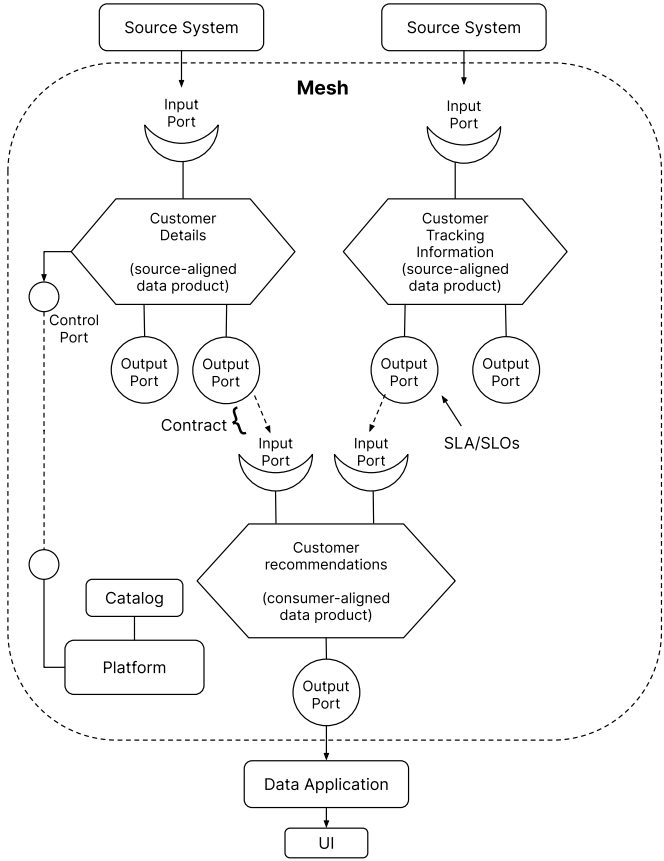

Fig. 1. A small example data mesh consisting of three data products

In the example illustrated in Fig. 1, this one output port is consumed by a data application, i.e. a use-case implementation with a dedicated user interface. Similar to the source data systems, this data application resides outside of the data mesh because it is optimized for end-user-interaction (human or machine). It is not optimized for composition, i.e., it is not designed in such a way that another data application can be built on top of it by combining it with other data sources. All data products, on the other hand, should be designed for composition and therefore reside inside the mesh.

In fact, a large part of the innovation potential of data mesh lies in the network effect that stems from the ability to quickly create new data products by combining data from existing data products. Therefore, one important goal of any data mesh implementation and its infrastructure platform should be to enable simple and reliable composition.

In our experience, the right approach to input/output port linkage is key to reliable data product composition. Data mesh takes a lot of inspiration from site reliability engineering. This is particularly visible in the adoption of service level objectives (SLOs) and service level agreements (SLAs) for data products. Because SLOs are often highly technical and therefore depend on the specific way the data is made available, it makes more sense to attach SLOs to individual output ports than to attach them to a data product as a whole. An input port on the other hand should state its assumptions and requirements. One way to do this is in the form of an executable contract test. In the consumer-driven contract testing approach, such an executable contract test is sent to the data provider (the output port). There, it is executed against the actual data that is provided by the port, so that an alert is triggered if the data does not meet the codified expectations.

With such clearly defined expectations and assumptions from both sides, reliable data product composition can be achieved, which is key for a data mesh that can scale. A data mesh platform should provide as much automation support as possible for data product composition. For example, to enable data product discovery through a central data catalog, data products can interact with the platform through control ports as indicated in Fig. 1.

### B. A Data Product Model with Port-Level Composition

Now that we have clarified what ports are for, we can describe how the concept of input/output ports relates to other data mesh concepts, which is illustrated in Figure 2. This conceptual model is meant as an extension or refinement to what is called a domain model in [5]. In this extension, we focus solely on technical aspects of data products and their composition. We are well aware that organizational principles are a major part of what constitutes the data mesh approach. For instance, any data product should be owned by exactly one data domain, and in data mesh, such a data domain consists of people – the respective domain experts. These organizational concepts are, however, beyond the scope of this paper.

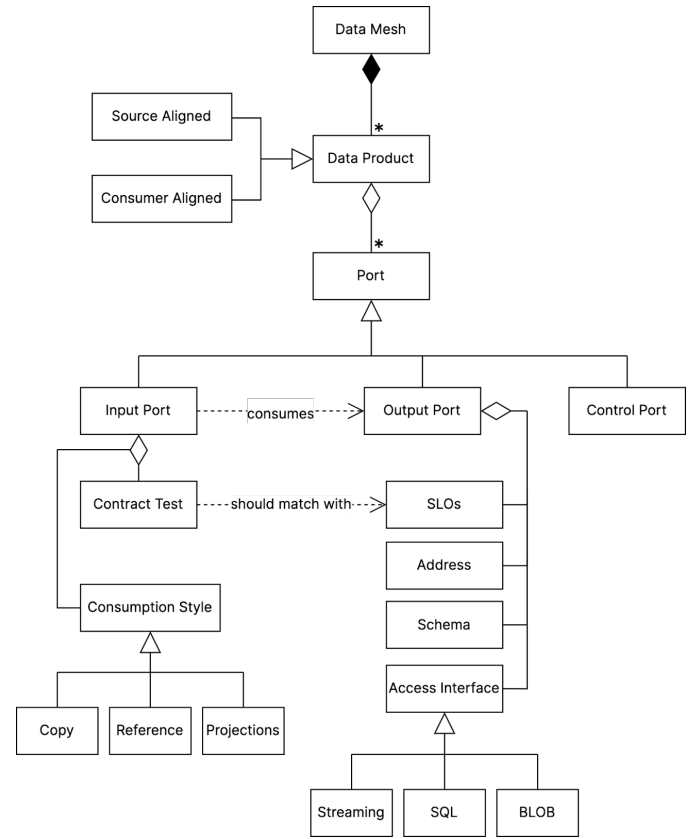

Fig. 2. A data mesh conceptual model focusing on the port concept.

As mentioned above, a data mesh – from a technical perspective – consists of data products, which make up the nodes of the mesh. Each data product (be it source-aligned or consumer-aligned) has several ports, either input or output. An output port has different properties than an input port. An output port should explicitly state its SLOs. To be programmatically consumable, an output port needs to have an address and a schema. Furthermore, it needs to state its access interface type, which can be either BLOB storage, streaming, or SQL-based. The notion that an output port must have exactly one of those access interface types is one of the reasons why data products have multiple output ports – one port for each required access type.

An input port must reference exactly one target output port from which it consumes data. It should state its assumptions w.r.t. to this output port, ideally in some automatable form such as a contract test. This contract test should match with the SLOs published by the referenced output port. Furthermore, an input port must implement one of three consumption styles: by copy, by reference, and by projection. This consumption style determines whether data is actually copied to data infrastructure owned by the consuming data product or physically stays at the output port's underlying data infrastructure. The control port is the data product's interface to the data mesh platform, e.g. to publish its metadata to a central data product catalog.

## III. Components of a Data Mesh Platform

The purpose of a data mesh platform is to make building complex data solutions as easy as possible. It aims to reduce friction by automating processes to make it easier for teams to exchange data with each other. This ideally leads to faster business value and reduction in duplicate efforts.

The primary goal of a data mesh platform is to allow for scaling the decentralized creation of data products across an organization by avoiding duplication of infrastructure efforts and thereby reducing the need for data infrastructure skills in every data product team. Thus, the main objective of the platform is to make data product creation and maintenance as easy as possible. This is the purpose of the data product developer experience plane, that we mentioned in Sect. I.

The other important objective of a data mesh platform is to provide a set of centralized services that are needed to enable the desired network effect of data product composition. This is the purpose of what is often called a *mesh experience plane*. An important service provided by this plane is a central data product catalog that allows users to search for existing data products. Creation and management of global standards and policies are other aspects of this plane.

To make sure an organization's data is following such standards and policies, the platform needs to be able to enforce standardization at various levels. At the same time, a data mesh platform also needs to provide flexibility. A common scenario in a large organization is that a domain team needs to embed their existing business processes into their data products, which can be challenging if the platform does not support customization of its services. The platform should therefore allow data product teams to extend existing features as long as they respect the contracts offered by the platform.

Data mesh focuses on distributing the responsibility of managing the complexity of data across multiple domain teams within an organization. This stands in contrast to traditional approaches that force a centralization of data to control its complexity. Data mesh's decentralization, however, creates a risk of forming silos and of creating incompatibility between data produced by different teams. A typical challenge in a data mesh is thus the right balance between the two contrasting ideas of federated decentralization and common standardization of data processes and practices.

### A. Balancing Decentralization with Standardization

On the one hand, a data mesh platform needs to enable decentralized domain teams in such a way that they can build and evolve their data products without having to check back with a central authority. This decentralization, which allows domain teams to act independently, is key for the mesh's ability to scale. On the other hand, the platform also plays the role of a standards-promoting mechanism that counteracts the heterogenization that stems from the domain teams' independence. In data mesh, however, standardization is not enforced in a top down fashion but instead by support and convenience: If the tools provided by the platform are easy to use and help the domain teams to achieve their goals more efficiently, then these tools will be used out of convenience. This tool usage will then shape the way domain teams build data products and therefore drives homogenisation on an implementation level. This is even more effective if the tool usage allows the central data mesh platform to keep a certain amount of control over data products that run on its infrastructure. For instance, when a platform-provided template has been used to create a data product, this data product should be updatable to reflect the newest changes on this template.

This two way communication between data products and the data mesh platform can be achieved by what is commonly referred to as a *sidecar*. A sidecar is a library or process which is part of each data product but is maintained by the platform. There are two approaches to this: In the more decentralized approach the sidecar is a library or application that *runs* as part of the data product but is maintained and distributed by the platform. In the less decentralized approach a sidecar is a process that only logically belongs to a data product but runs as part of the platform. There are pros and cons to each of these approaches. In the more decentralized sidecar approach, the platform needs to maintain multiple instances of the library or application which is providing the functionality to each of the data products. At the same time, this approach does not create a central bottleneck. In the less decentralized approach, there are less instances for the platform to maintain. However, care has to be taken to build the platform in a scalable manner so that it does not create a bottleneck. In our experience it is a good idea to start with the more centralized approach using technologies that support distributed scaling and move towards a more decentralized approach if needed.

To summarize these sometimes conflicting requirements of a data mesh platform: An ideal data mesh platform is all about enabling logical decentralization while at the same time being a highly centralized piece of infrastructure that should strive for keeping as much control over the specific technical implementation as possible.

## B. Example Platform Architecture

Based on the concepts and requirements presented so far, we have created a reference implementation of a data mesh platform at our organization. In the following, we present an example architecture that we derived from our implementation. This architecture illustrates how a data mesh platform might look like and can be used for guidance when building a data mesh platform. Of course, there might be many other good approaches to designing a data mesh platform.

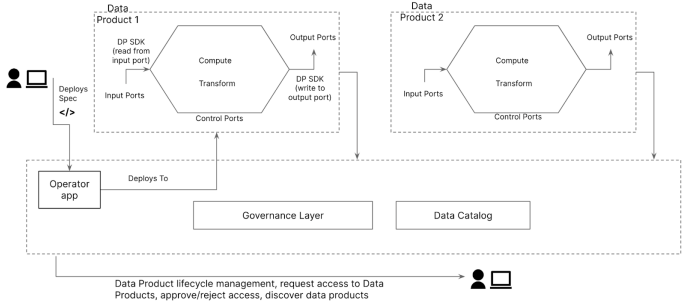

Fig. 3. An example architecture of a data mesh platform based on the data mesh reference implementation we built within our organization.

As shown in Fig. 3, the core concept of our architecture is that of a data product. A data product interacts with its environment (the mesh) via ports. Besides data products themselves, there is a set of data product lifecycle management tasks that need to be accomplished. In our architecture these capabilities are provided by an *operator app*, a *data catalog*, and a *governance layer*. The operator app helps to orchestrate the various tools in the ecosystem and manages the lifecycle of the data products. A data catalog enables data discoverability and lineage. The governance layer is a set of tools which manage governance rules such as access control, both at the mesh level and at the data product level. In this particular architecture, the operator app provides APIs for both the data product developer experience plane and the mesh experience plane. The tools in the governance layer and data catalog help with discoverability of data products and with maintenance of business rules and policies.

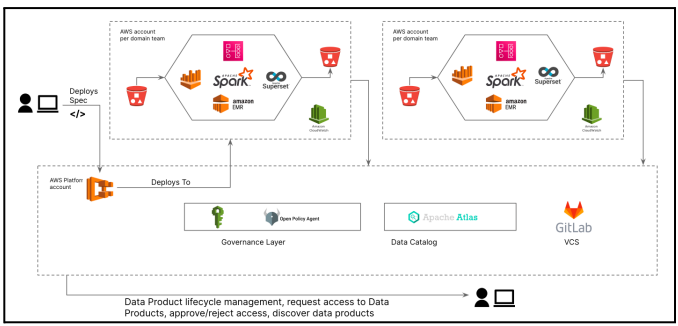

Fig. 4. The technologies used in the data mesh reference implementation

Fig. 4 shows the specific technologies that we use in our reference implementation. AWS is used as the cloud provider and its services are used for most of the data product related resources – S3 for batch ports, Managed Airflow as the current de facto standard for data pipeline orchestration, EMR for executing data pipelines, and Athena for querying. For monitoring and visualization, we use Grafana and Apache Superset, respectively. The roles and access policies are managed using IAM. For the data catalog, we use Apache Atlas. Gitlab is used as the version control solution.

## IV. Data Governance as Part of the Platform

Now that we have discussed the general structure and requirements of a data mesh platform, let's look at how the specific requirements of federated data governance affect platform design. Data governance, in general, is a field of activities that aim to ensure the trustworthiness of data and to minimize data related risks [20]. Examples of such activities are data security audits or the definition of company-wide data quality standards. Overall, data governance encompasses a broad variety of topics such as data quality, visbility, ownership, consistency, security, access control, and compliance. The first topic that has to be tackled, however, is usually that of data access control. Data mesh's fourth principle "federated computational data governance" describes a decentralized approach to data governance that is enabled by automation provided by the data mesh platform. Therefore, we will focus our discussion of data governance on how to achieve decentralized data access control.

Decentralized access control means that the decentralized domain teams who are responsible for their data products are the ones who decide who can consume their data products. This stands in stark contrast to the typical situation where for instance a data protection officer and their team fulfill this task from a central position. This is usually argued necessary and not decentralizable because of liability and legal issues. A data mesh platform therefore needs to provide the means to distribute individual access decisions to domain teams. At the same time it has to allow for central functions, such as a data protection officer, to prescribe, enforce and review global access control policies. An example of this could be the following: Access to critical financial information can only be granted to people who are listed as insiders at a country's financial trading supervision authority. This is an example of a global policy that can be put in place by a central function. An individual data access request, however, should then be handled by the requested data product's team.

This means that the platform has to enable domain teams to make those decisions locally while making sure that global policies are adhered to, e.g., by disallowing any data access grants to persons or services whose roles are not sufficiently specified. Furthermore, the platform has to provide the means to tag which data is classified as critical financial information. The tagging itself, however, should be done decentralized by the respective domain team, as we will see in the next section.

### A. Platform Requirements of Decentralized Governance

In this section we discuss how a data mesh platform can support the interplay of global policies and decentralized decision making, that we described before. In order to infer the use-cases that a platform needs to support, we present a slightly more involved example of the governance challenges in a decentralized setup.

Let's imagine a source-aligned data product $DP^A$ that is serving web tracking data about a company's users. This data contains personal identifiable information (PII) and therefore is considered highly sensitive. The first two activities around governance which the platform has to support are the following.

First, there needs to be a way for data protection experts to define global data sensitivity levels as well as policies on how data of each level needs to be handled in order to stay compliant with government regulations and to protect the rights of its customers. An example of three sensitivity levels and their policies could be (a) "financial data" which means that access needs to be restricted to persons on the insider list, (b) "highly sensitive" which means that the data needs to be always stored encrypted and (c) "sensitive with PII". In this case it has to be both encrypted and needs to exhibit a traceable reference to the affected person in order to be able to delete all data about a specific person in case that person files a data deletion request according to the EU General Data Protection Regulation's (GDPR) "right to be forgotten".

Second, the decentralized domain team that is responsible for the source-aligned data product has to be provided with tools to easily tag their data, applying those globally defined sensitivity levels. This tagging functionality has to be provided in such a way that the platform can automate the implementation, or at least check the fulfillment of the policy associated with the respective sensitivity level. This way, we avoid that the central governance team becomes a bottleneck because they are busy with trying to understand the domain details of each data product. One way of forcing domain teams to at least use the global sensitivity levels, is to refuse the processing of any data that is not tagged on the shared infrastructure which the platform is providing.

How is the data product composition that we discussed in Sect. II connected to decentralized data governance? Decentralized governance will not scale if every data product team that is consuming data from another data product has to re-decide how sensitive the consumed data is. This would require intricate domain knowledge of the consumed data's domain. Instead, data sensitivity tags have to be carried over transitively along higher-order data product composition, and this has to be actively supported by the platform.

Let's stick with our example of a source-aligned data product $DP^A$ that serves PII-containing data. Here, the team is using the global sensitivity levels to tag their data accordingly. This means that any output port of the data product which is serving such data, consequently states that its data is PII-sensitive. If another data product $DP^B$ consumes this data via one of its input ports, all of the output ports of $DP^B$ should be marked as PII-sensitive by default. Platform support for this means that if a new composed data product is created using platform tools, then this transitive tagging of output ports will happen automatically. One of the tools that bakes in these principles of metadata propagation through lineage is Apache Atlas [14]. This is one of the reasons why we chose to use it as part of our reference platform implementation.

Of course, the team responsible for $DP^B$ should not be hindered in their autonomy and needs to be enabled to make local decisions applying their domain expertise. For example, if one of their output ports is only serving data that has been cleansed from any PII, then the team has to be able to override the automatically assigned data sensitivity level. Vice versa, if a data product is only consuming non-sensitive data, but is combining it in such a way that it becomes sensitive in some way (e.g. sales data that allows for a revenue forecast), then the team also needs to be able to override the automatically assigned sensitivity level. Fig. 5 illustrates this interplay of global policies, localized decisions, and platform support.

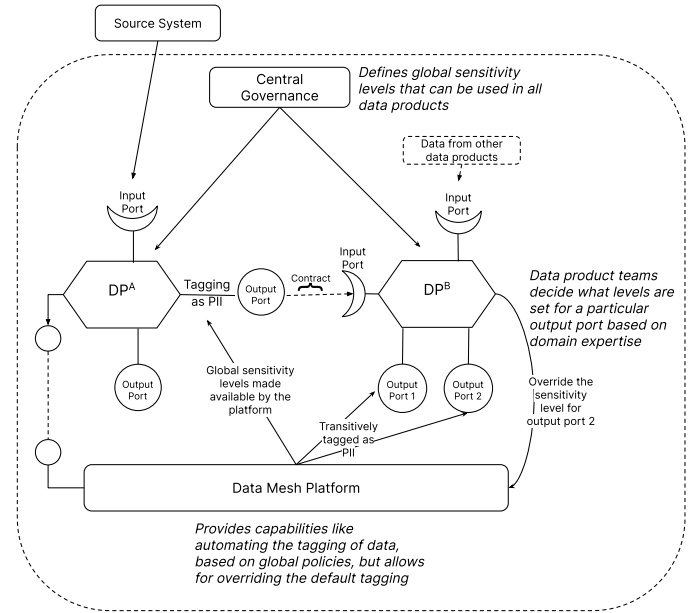

Fig. 5. Interplay of global policies, localized decisions, and platform support by the example of data sensitivity levels.

In order to not render all central governance ineffective because everyone can simply override their data sensitivity tags, such overrides in higher-order data products need to be reviewed and approved by central governance. This, of course, entails the risk of central governance becoming a bottleneck again. Therefore, this approval process needs to be automated as much as possible. For instance, any override that applies a sensitivity level that is more strict than the one of the input data could be approved automatically. Similarly, on a data product level, as many data access requests as possible should be handled automatically by role-based authorization (RBAC), so that a data product team only needs to specify what kind of data they serve. The actual authorization should then be governed by global role-based policies that are only informed by local domain knowledge.

### B. Implementation Approaches

When discussing means of access control, there are three popular contending approaches. All of them implement the general idea of *computational governance*. This means that there is a central authority that, upon getting a request, decides whether to grant or deny access by executing automated rules on the meta properties of the data and consumer. The request may be granted as an execution of a single rule or through a collective context of many different rules and access control lists (ACL).

As an example, consider the scenario where data products A and B are two data products in a *marketing domain*. A global default rule of the platform is to deny all access requests. However, a domain-specific access rule defined by the marketing domain allows all data products within the marketing domain to access each other's output ports. In this scenario data product B can access all data from data product A since the domain-specific ACL overrides the global policy.

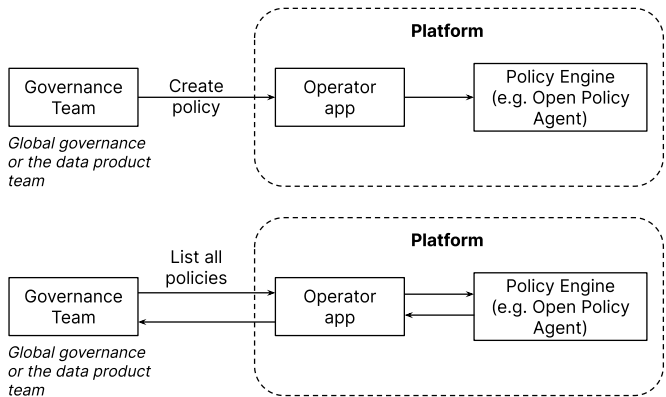

Fig. 6.  Managing policies, using Open Policy Agent as the policy engine

Fig. 6 presents a simplified version of how we approach policy management in our reference platform implementation. We are using Open Policy Agent (OPA) as the policy engine which can be used for policy authoring as well as policy enforcement. It provides a high-level declarative language that can be used to specify policies as code. Now that we have seen how rules may be applied and overridden, let's dive deeper into their method of execution and the different approaches available.

*1) Gateway Approach:* One of the more commonly implemented approaches is that of a central governing gateway server that processes all incoming requests and acts as a centralized query and access server very similar to typical OLTP (online transaction processing) access-authorization patterns. A governance steward or data product developer can describe the RBAC policies for their specific datasets and schemas which are enforced when a client or user connects via their credentials. This is often a system user that authenticates via JDBC with the gateway server which may or may not process the results itself. In case of most tools with separate storage and compute (e.g., Spark, Presto, Impala) the authorization gateway acts as a JDBC proxy which parses the query, checks if the credentials for the session are valid, and then forwards the request to a massively parallel processing tool which largely processes the data itself and hands back the result. The advantage of this approach is that it works seamlessly with any client since JDBC based authorization has been around for a long time. Also, this approach (illustrated in Fig. 7) offers powerful attribute based access control at a column level. However, it may fail to scale with very large datasets, large numbers of customers, and or very frequent system queries. It also mostly works with relational databases and datasets.

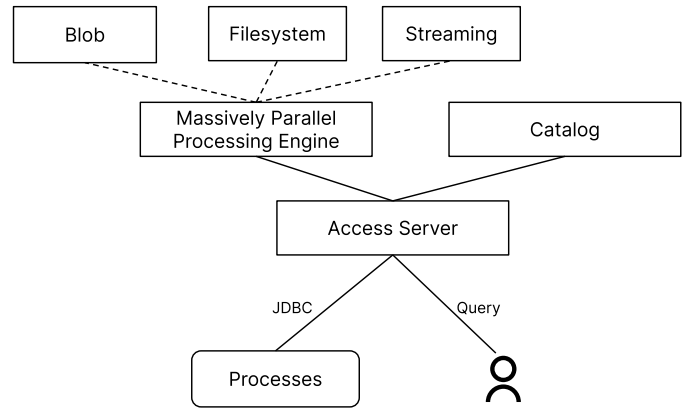

Fig. 7.  Gateway approach with a central data access server – examples are Presto, RBAC, OLTP

*2) Token Approach:* Another approach that is gaining popularity is that of a token generator or proxy server. Leveraging the power of decoupled compute and storage processing systems like Apache Spark, this approach provides the added benefit of scaling per client without becoming a bottleneck. The proxy access server, instead of providing the data directly, authorizes the incoming request for datasets, and fetches meta information from the catalog. In case the source query successfully passes the check, it returns a token to the underlying system as well as information on how to authenticate directly with the underlying storage. We illustrate this approach in Fig. 8.

Such systems not only scale well but also work with non-relational datasets like images, BLOBs etc. The only downside is that they may not work well with all data storage types or need custom implementation for the underlying storage. Databricks Delta Sharing [10] is an example of a framework based on this approach.

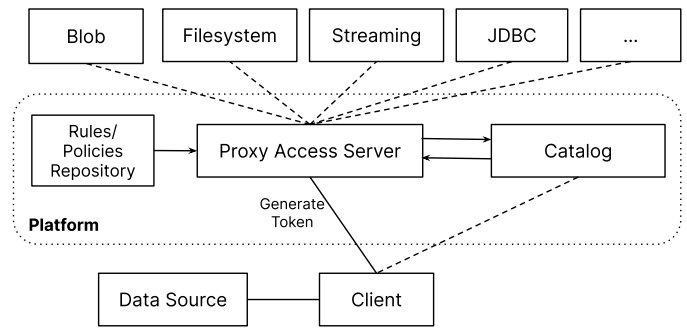

Fig. 8.  Token approach – examples are Apache Ranger, Privacera

*3) Encryption At Rest Approach:* In this pattern of authorization enforcement, data is encrypted at rest by the producer of the data using an encryption key. A central key management server (KMS) hands out decryption keys to the client. There aren't a lot of tools making use of this approach because of the additional compute time spent encrypting and decrypting the data many times. However, a big advantage of this approach (which we illustrate in Fig. 9) is the additional security it brings. Even if a system copies data from the source system, it needs the key as well as the compute to decrypt the data that is stored. Data formats like parquet are good storage formats for such an encryption scheme [11].

For example, when dealing with personal health information and other forms of sensitive health and personal data, this is often a matter of compliance and approaches like homomorphic encryption are important to provide an identity preserving way of processing data [13].

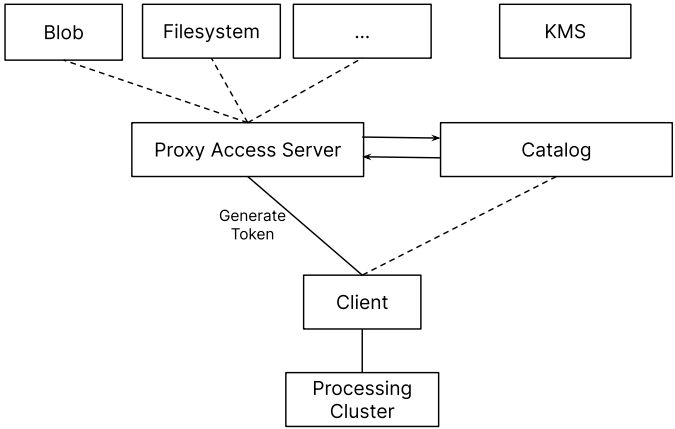

Fig. 9. Encryption at rest approach which requires custom tooling.

Another alternative to the above three approaches that we would like to mention is where the access server upon creation of the policy translates or maps the policy to that of the underlying compute, e.g., S3 bucket policy, database role and restriction. Some Apache Ranger plugins for cloud based storage commonly use this approach to enforce authorization natively on cloud resources [12].

An important part of making any of these approaches work in practice is to ensure that enough frictionless abstractions exist, so that the user never needs to know how the underlying access is being granted. In the current technology landscape, no single tool covers all aspects required for implementing federated computational governance. In our experience, a combination of existing tools and standards employing a hybrid of these approaches need to be applied. On top of that an abstraction needs to be created, so the users in the ecosystem can use them easily. For example, the gateway server approach works well in the beginning when no systems exist, but should be replaced with a token based approach for large or frequently queried datasets. Finally, when protection of sensitive data is required, we suggest employing the encryption-at-rest scheme to protect copies of the sensitive data as well. The abstraction also allows the platform to evolve the underlying tooling with minimal effects on the user.

## V. Related Work

Existing work related to the one presented here can be divided into two categories: (a) existing academic work, and (b) existing technologies and frameworks for implementing a data mesh platform.

In the academic space, there is little related work in general, mostly because of the relative youth of the topic. Goedegebuure et al. recently performed a systematic literature review on data mesh in [16] and list only four published academic papers, of which three are from the same group of authors. That group presented the data mesh conceptual model [5], which we extended in Sect. II. While the presented model is very useful to advance the discussion about data mesh in general, we feel that it lacks the technical depth to help building a scalable data mesh. Specifically, the concept of input and output ports to enable data product composition is only touched upon. This is why we focus on this aspect in our extension of their conceptual model. From the same authors, in [6] a technical architecture is presented but it is only demonstrated with a very limited proof-of-concept (PoC). Similarly to our reference platform, Apache Atlas is used for data product discovery. However, the PoC presented also seems to lack the concept of multiple input and output ports per data product and does not use public cloud infrastructure. The only other published academic paper is that of Butte et al., who present a generic data mesh architecture [21]. Similar to our reference implementation, the architecture is based on AWS infrastructure and input and output ports are identified as key concepts for data product composition. However, the paper does not go into much further detail about the implementation.

In the Saxo Bank case study [7] our colleagues from Thoughtworks already described a specific industry example of how to approach data governance in data mesh. However, its relation to the data mesh platform is not covered in much detail. An example of a case study from the pharmaceutical industry is the data mesh adoption journey of Roche [8]. The architecture and platform pieces that our colleagues created at Roche have a large overlap with the reference implementation that we present here.

With regards to existing technologies and frameworks, one of the shortcomings we see is that they often focus on only a few aspects of data mesh, such as governance, while missing out on how to tie all the different aspects together. This becomes apparent, for instance, in the frequent lack of a dedicated data product developer experience plane or limited efforts to create an architecture that can be extended by developers. An example of such a useful but somehow one-sided implementation which uses AWS Lake Formation and AWS Glue has been presented in [3]. In fact, when starting to design our reference implementation, we evaluated the proposed solution and found it missing certain key aspects. For example, it is missing a developer experience plane, the "language" of data mesh (i.e., input/output ports and data products), and the ability for users to customize and replace parts as they see fit. Many offered solutions lack this flexibility due to a strong tie-in to their own ecosystems. Therefore, they do not allow for integration with other open source tooling for accomplishing the various aspects of data mesh. An example of an – in our opinion – rather closed data mesh platform implementation was presented in [9].

## VI. Conclusion & Future Work

Achieving effective decentralized data governance imposes a whole set of conceptual and practical challenges. With this paper, we aim to ease some of these challenges by connecting general concepts of data mesh with specific data mesh platform concepts and data governance concepts in a cohesive way and by sharing our practical experiences with building a data mesh platform with data governance support. We therefore presented an extension of an existing conceptual model of key data mesh concepts and a set of approaches on how to achieve federated computational data governance through data mesh platform means.

Looking at the current state of the industry, we find that while there is a certain maturity on what data mesh is and how it can help organizations, there is not much guidance on how to map the general concepts of data mesh to specific tools and approaches. An approach on how to make existing storage and compute compatible with policy based access control mechanisms has recently been described in [23], which we find is a good starting point for maturing the technology needed for decentralized data governance. However, more research is needed on how the organizational aspects of data mesh work can be accompanied by the right technical approaches. There is also some potential in consolidation and standardization of approaches across the industry. For instance, a standardized data product description language similar to the web services description language (WSDL) could accelerate the first steps of data mesh adoption within an organization and even allow for cross-organization data product sharing. We are therefore exploring whether we can extract the data product description format used in our reference implementation in order to use it as a foundation for such a standardized data product description language.

We hope the concepts and approaches presented in this paper can help with data mesh platform design and can serve as a guideline on how to interpret data mesh concepts and pave the way for future research.

ACKNOWLEDGMENT

Many people have contributed to the discussions and the concepts leading up to the ideas presented in this paper. We especially would like to thank our colleagues from Thoughtworks, the reference implementation team, and the Data and AI leadership team, in particular Vanya Seth. Special thanks goes to Zhamak Dehghani for creating data mesh in the first place, for relentlessly educating about it, and for the feedback we received from her on our reference platform implementation. Arif is grateful to "IFAF–Institut für angewandte Forschung Berlin e.V." for funding parts of this research. Finally, we want to thank the anonymous reviewers for their valuable feedback.